\documentclass{osa-article}

\journal{oe}
\usepackage{siunitx}
\begin{document}

\title{High-frequency broadband laser phase noise cancellation using a delay line}

\author{Michał Parniak,\authormark{1,2,*,**} Ivan Galinskiy,\authormark{1,*}  Timo Zwettler,\authormark{1,3} and Eugene S. Polzik\authormark{1}}
%consider authors
\address{\authormark{1}Niels Bohr Institute, University of Copenhagen, Blegdamsvej 17, 2100 Copenhagen, Denmark\\
	\authormark{2}Centre for Quantum Optical Technologies, Centre of New Technologies, University of Warsaw, Banacha 2c, 02-097 Warsaw. Poland\\
	\authormark{3}Present address: Institute of Physics, École Polytechnique Fédérale de Lausanne, 1015 Lausanne, Switzerland\\
	\authormark{*}These authors contributed equally to this work\\}

\email{\authormark{**}parniak@nbi.ku.dk}

\begin{abstract}
		Laser phase noise remains a limiting factor in many experimental settings, including metrology, time-keeping, as well as quantum optics. Hitherto this issue was addressed at low frequencies, ranging from well below 1 Hz to maximally \SI{100}{\kilo\hertz}. However, a wide range of experiments, such as, e.g.,  those involving nanomechanical membrane resonators, are highly sensitive
to noise at higher frequencies in the range of \SI{100}{\kilo \hertz} to \SI{10}{\mega\hertz}, such as nanomechanical membrane resonators. Here we employ a fiber-loop delay line interferometer optimized to cancel laser phase noise at frequencies around \SI{1.5}{\mega\hertz}. We achieve noise reduction in \SI{300}{\kilo\hertz}-wide bands with a peak reduction of more than \SI{10}{\deci\bel} at desired frequencies, reaching phase noise of less than \SI{-160}{\deci\bel(\radian^2/\hertz)} with a Ti:Al$_2$O$_3$ laser. These results provide a convenient noise reduction technique to achieve deep ground-state cooling of mechanical motion.
\end{abstract}

\section{Introduction}
Lasers are nowadays well-established as the workhorse of modern telecommunication, metrology, as well as developing quantum technologies. However, as has been already realized by Schawlow and Townes \cite{PhysRev.112.1940}, the phase noise of a laser is finite and fundamentally limited. Practically, many more effects, such as thermal \cite{NUMATA2012798} or acoustical \cite{CONFORTI2020126286} noise of laser cavities, contribute to the total phase noise. 
Many techniques are employed to reduce phase noise, and thus the linewidth of lasers. To great success, external ultra-stable cavities have been employed as references  \cite{Kessler2012,PhysRevA.77.053809}, achieving linewidths well below \SI{1}{\Hz}, but broadband phase noise remains a problem in many applications.  In contrast to intensity noise, which usually exhibits localized peaks due to relaxation-oscillation \cite{8123620}, phase noise in lasers tends to exhibit both a broadband noise floor and technical noise peaks.

The broadband noise can be either suppressed passively via filtration through a narrow cavity, or using an active feedback signal. As an alternative method to reference cavities, an unbalanced Mach-Zehnder interferometer with an optical fiber delay line in one of its arms can be used. Such setups are routinely employed to characterize laser linewidths  \cite{LUDVIGSEN1998180}. One can then use the signal of such a delay-line setup to feedback on the laser light's phase, thereby reducing its noise \cite{Li:17,Chen:89,Cranch:02,Sheard:06,Kefelian:09,Dong:15,  _m_d_2015,Shehzad:19,doi:10.1063/1.3606439,Bandutunga:20}. All previous approaches operated in a relatively low-frequency range, starting from very low infrasonic frequencies up to maximally around \SI{100}{\kHz}. This has so far satisfied most needs of spectroscopy experiments, and implementing feedback at higher frequencies becomes challenging. 

 For a delay-line setup, fiber noise becomes an important matter to consider. At low frequencies, acoustic isolation of fibers has been successfully implemented \cite{Bandutunga:20}, but at higher frequencies the thermomechanical and thermoconductive noises are unavoidable \cite{PhysRevA.86.023817}. Fiber noise also poses limitations on transfer of optical frequency standards, and thus a reverse approach is often employed in which the fiber is stabilized to a narrowband stable laser \cite{Ma:94,PhysRevLett.111.110801}. Several works have also studied fundamental limits to fiber noise and related fiber strain sensing \cite{doi:10.1063/1.4939918,Gagliardi1081,Cranch286}.

Here we employ a 50-meter-long fiber delay line combined with a balanced detection scheme to measure laser phase noise and subsequently use active feedback to reduce it at high frequencies. In particular we are interested in reducing phase noise at frequencies in the vicinity of \SI{1.5}{\MHz} that correspond to the resonance frequency of a membrane mechanical oscillator \cite{Tsaturyan2017,ourOptica}. Using a feedback loop, with the gain concentrated around the frequency of interest, we achieve phase noise of \SI{-164}{\deci\bel(\radian^2/\hertz)} at 1.5 MHz frequency offset of a Titanium-Sapphire laser (M-Squared SolsTiS), providing noise reduction in a previously untackled frequency regime with a very high absolute bandwidth of around 300 kHz, compared with hitherto approaches that target up to 100 kHz bandwidth at baseband. The results are enabled by low-noise detection at high light powers, combined with an optimized fiber length and properly engineered feedback.

Noise at such high frequency offsets is an essential limitation in quantum optomechanics, where the sideband at the resonance frequency of the oscillator leads to significant classical drive forces \cite{PhysRevLett.123.153601,Kippenberg_2013,Safavi_Naeini_2013}. We envisage that using this light for feedback cooling of membrane resonators \cite{Tsaturyan2017,Rossi2018,ourOptica} in a sideband-resolved cavity quantum optomechanics regime will allow to break an important barrier of 0.1 residual occupation of phonons of a membrane resonator mode at liquid helium temperatures ($\sim \SI{4}{\kelvin}$).

The paper is organized as follows. We first introduce the experimental setup, which involves two similar unbalanced interferometers, for feedback and characterization. Next, we introduce the model for detection of the laser's phase noise, including a treatment of spurious fiber noise and photon shot noise, as well as a model for feedback. Finally, we show results for noise suppression in several MHz-level bands and compare them with our model prediction.

\section{Setup}
\label{sec:setup}
\begin{figure}[h!]
	\centering\includegraphics[width=1\linewidth]{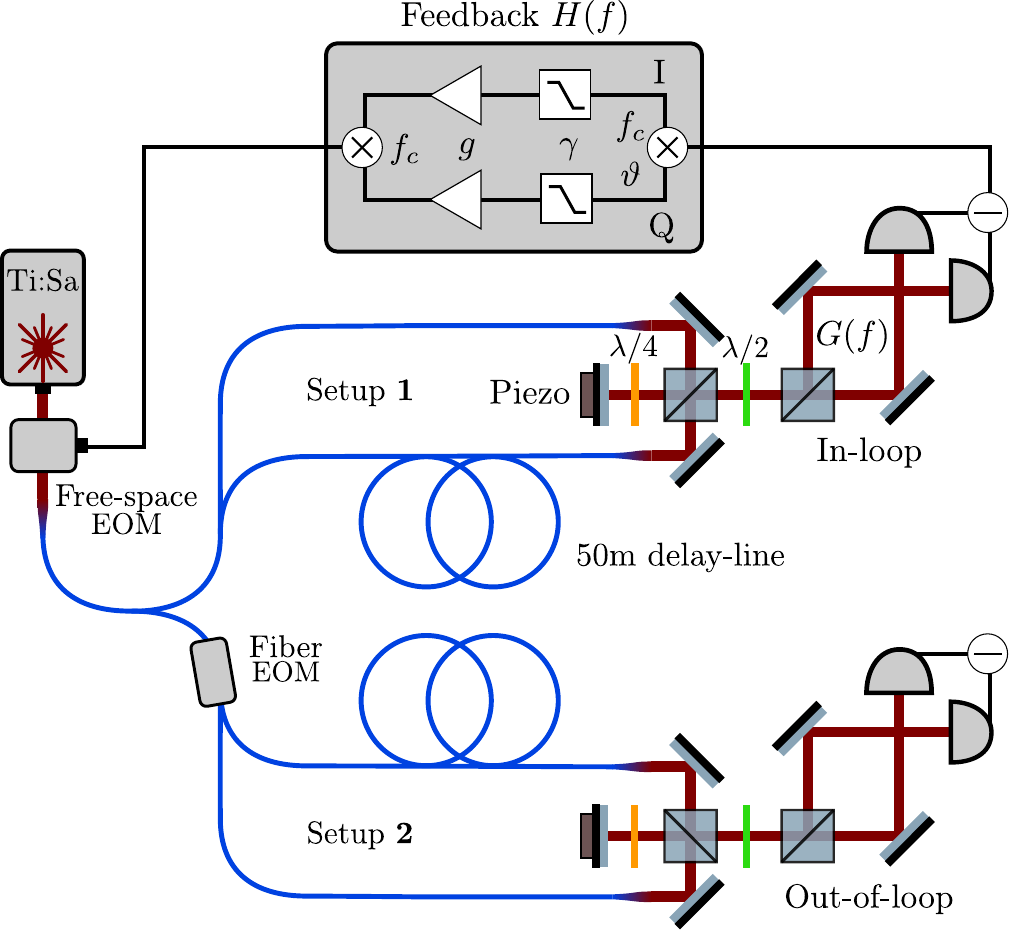}
	\caption{Schematic of the delay line feedback loop (setup 1) and an additional loop for out-of-loop measurement (setup 2). At the input, the interferometers feature a fiber polarization-maintaing 50:50 beamsplitter, while at the output the beams are combined via free-space polarization optics and sent onto a balanced photodetector. Rotating the half-wave plate ($\lambda/2$) allows fine adjustement of balanace on the detector. Both interferometers are independently stabilized with piezo mirrors at low frequencies, such that the detection remains balanced. Feedback to the laser light is applied from setup 1 via an electronic controller through a free-space EOM. An additional fiber EOM is used for absolute calibration of setup 2.}
	\label{fig:exp_schematic}
	
\end{figure}
As presented in Fig. \ref{fig:exp_schematic}, our test experiment consists of two unbalanced Mach-Zehnder interferometers with the short arm being $~\SI{1}{\meter}$-long and the long arm being  \SI{50}{\meter}-long. All fiber in our setup is polarization maintaining (PM), single-mode fiber (PM-780HP), which allows us to maintain good interferometric visibility and polarization stability. We inject equal amounts of light into both arms using a 50:50 fiber beamsplitter. In each interferometer we place an additional piezo-actuated mirror in one of the arms for stabilization of low-frequency phase drifts in order to lock the interferometer at the optimal (balanced) point for sensing phase noise. The beams are combined on a polarizing beamsplitter (PBS), and then sent through an additional half-wave plate and another PBS onto a balanced detector.

Our custom detectors are based on two Hamamatsu S5971 silicon photodiodes connected in series in a differential configuration \cite{TheJurgen}. The photodiodes themselves feature a high quantum efficiency of $\eta\approx0.9$, which is combined with separate amplification paths for DC and AC ($> \SI{150}{\kHz}$) components of the signal. The overall transimpedance gains are \SI{130}{\kV/\A} and \SI{30}{\kV/\A} for AC/DC respectively. The low-frequency DC output is used to actively stabilize the interferometer with the piezo-actuated mirror. This ensures that the detector always operates in the balanced regime, rejecting the amplitude quadrature of the incoming light to better than \SI{20}{\dB} and preventing the amplifiers from saturating. Due to large optical powers used, we maximize the size of the beam at the photodiodes to prevent temporary "bleaching" of the photodiodes by excessive intensity.

In setup 1, the high-frequency AC output is sent to a STEMlab Red Pitaya 125-14 board equipped with PyRPL software \cite{8087380}, which implements the feedback filter in a field-programmable gate array (FPGA), and drives a free-space EOM (electro-optic modulator) through an additional \SI{20}{\dB} attenuator to minimize electronic and quantization noises. The signal is also sent to an FFT spectrum analyzer in order to measure in-loop noise.

The feedback is achieved using an I/Q modulator/demodulator module in the RedPitaya/PyRPL FPGA architecture. We first demodulate the error signal with a carrier wave at a desired frequency $f_\mathrm{c}$. Subsequently, we apply a lowpass filter with a bandwidth $\gamma$ independently to both quadratures and multiply both signals by gain factor $g$. Finally, the signal is modulated with the same carrier wave $f_\mathrm{c}$ shifted in phase by $\vartheta$. Using the same carrier wave at demodulation and modulation leads to cancellation of unwanted carrier phase noise. Stability is guaranteed since the entire scheme is implemented digitally in the FPGA. The phase difference of the carrier wave between demodulator and modulator determines the phase $\vartheta$ of our feedback, which is empirically optimized. Overall, the filter is shaped like a Lorentzian centered around $f_\mathrm{c}$ with a bandwidth $\gamma$. Importantly, it allows us to continuously tune the phase of the feedback at $f_c$, thus compensating the phase lag introduced due to a delay.

In setup 2, we send the AC signal to a spectrum analyzer for out-of-loop noise measurement. It uses an additional fiber EOM which can be used for absolute phase-noise calibration as it features a flat, calibrated response to phase modulation \cite{GorodetksyDOVOCRUFNC}.

Our laser operates at $\lambda=\SI{852}{\nano\meter}$ and outputs approximately \SI{200}{\mW} of light power. We send \SI{120}{\mW} through the free-space EOM for the entire experiment \cite{ourOptica}, part of which is sent to both fiber delay line setups.  In setup 1, we obtain $\bar{P}=\SI{11}{\mW}$ of total power impinging on the balanced detector, which allows for fiber-noise limited measurement of laser phase noise, increasing the signal-to-shot-noise ratio. This allows for the fiber noise to be larger than shot noise until up to 2 MHz. Furthermore, we minimize the beam path between setup 1 and the laser, which is less than 3 meters and includes only two mirrors.  For setup 2, we use $\bar{P}=\SI{5}{\mW}$ of total power at the detector. Setup 2 is connected using 5 meters of additional fiber. Any noise added in propagation will be treated as an additional detection noise.

\section{Model}
 \label{sec:model}
\subsection{Detection}
We consider input laser light with a fluctuating phase given by $\varphi(t)$. The phase of the field at the output of the long delay is given by $\varphi(t-\tau) + \varphi_f(t)$, where $\varphi(t-\tau)$ is the laser phase noise signal delayed due to propagation in the long arm by a constant time $\tau$ and $\varphi_f(t)$ is the noise introduced by the fiber. At the output of the short arm we simply have $\varphi(t)$. At the output of the interferometer we will measure the phase difference given by $\varphi(t)-\varphi(t-\tau)-\varphi_f(t)$.
The (frequency-domain) transfer function for the laser phase noise is therefore given by:
\begin{equation}
G(f)=1-e^{-2\pi i \tau f},
\end{equation}
with delay $\tau=nL/c$, introduced by the fiber of length $L$ with refractive index $n$ and the speed of light in vacuum $c$. 

The fiber noise $\varphi_f(t)$ is added independently as a detector noise term, together with shot noise. Active stabilization of the interferometer arms' relative phase preserves power balance, such that we can assume a total power $\bar{P}$ impinging on the balanced differential photodetector to be equally split between the individual diodes. In this configuration, the total power spectral density (PSD) of the registered optical signal is:
\begin{equation}
S=\eta \bar{P}^2 |G(f)|^2 S_{\varphi\varphi} + \eta \bar{P}^2 S_{\varphi_f\varphi_f}  + 2 h \nu \bar{P},
\end{equation}
where $\eta$ is the quantum efficiency of the detector.
The first term represents the laser phase noise and the second term represents fiber noise.
Finally, photon shot noise PSD is given by $2h\nu\bar{P}$ with $\nu=c/\lambda$ and $h$ being the Planck's constant.

Taking the transduction to/from phase noise into account, we may express the noise in terms of equivalent laser phase noise:
\begin{equation}
S_{\varphi\varphi}^\mathrm{meas} = \eta^{-1} \bar{P}^{-2} |G(f)|^{-2} S = S_{\varphi\varphi} + |G(f)|^{-2} S_{\varphi_f\varphi_f} +   |G(f)|^{-2} \frac{2 h \nu}{\eta\bar{P}}
\label{eq:phimeas}
\end{equation}
From this, we quite clearly observe that using a high optical power $\bar{P}$ diminishes the shot-noise contribution. The signal to noise ratio will indeed scale as $\sim1/(1+2h\nu/\eta\bar{P})$, therefore it is clearly beneficial to use as high power as realistically available.

For the sake of this model, let us assume that the two delay line setups (in-loop and out-of-loop) have identical powers and efficiencies. Conveniently, we can observe the laser phase noise directly in the cross spectral density (CSD) between the two setups:
\begin{equation}
S^{12} = \eta \bar{P}^2 |G(f)|^2 S_{\varphi\varphi}.
\label{eq:csd}
\end{equation}
This is possible as the laser phase noise is the only correlated noise shared between the two setups. In practice, we use this method to identify the laser phase noise contribution to total measured noise.

\subsection{Fiber noise}

In order to estimate the noise of the fiber, we use the theory of Duan from Ref. \cite{PhysRevA.86.023817}. The theoretical model leads to a conclusion that at our frequencies of interest the noise is heavily dominated by the thermoconductive noise, i.e. fluctuations of temperature transduced to phase fluctuations via thermal expansion and temperature dependence of the refractive index. The expected fiber noise is:
\begin{equation}
S_{\varphi_\mathrm{f}\varphi_\mathrm{f}} = \frac{2 k_B T^2 L }{\lambda^2 \kappa} \left(\frac{\mathrm{d}n}{\mathrm{d}T} + n \alpha\right) \frac{}{} \mathrm{Re}\left[\exp(2\pi i f r_0^2/2D) E_1(2\pi i f r_0^2/2D)\right],
\label{eq:fn}
\end{equation}
where $k_B$ is the Boltzmann constant, $T$ is temperature, $\kappa$ is thermal conductivity, $\frac{\mathrm{d}n}{\mathrm{d}T}$ is thermo-optic coefficient, $\alpha$ is coefficient of linear expansion, $D$ is thermal diffusivity,  $r_0$ is mode power profile radius and $E_1$ represents the exponential integral.
\subsection{Feedback}
We now introduce feedback, such that in the Fourier domain:
\begin{equation}
\check{\varphi} \rightarrow \check{\varphi} - H(f) G(f) \check{\varphi} - H(f) \check{\varphi}_f - H(f) \eta^{-1/2} \bar{P}^{-1} \check{\zeta},
\end{equation}
where $\check{\zeta}$ represents shot noise with PSD of $2h\nu\bar{P}$, while $\check{\varphi}$ and $\check{\varphi}_\mathrm{f}$ represent stochastic laser and fiber noises. The feedback transfer function $H(f)$ shall include both our desired feedback, as well as an undesired but unavoidable delay given by a prefactor $e^{-2\pi i \tau_\mathrm{D} f}$, mostly coming from the electronic processing delay ($\tau_\mathrm{D}\approx\SI{250}{\nano\second}$).
The feedback acts on light "shared" by both interferometers. In this case, setup 1 is the in-loop setup used for feedback, while setup 2 serves for independent out-of-loop characterization.

As a result of the feedback, the actual laser phase noise becomes:
\begin{equation}
S_{\varphi\varphi}^{\mathrm{fb}} = \frac{S_{\varphi\varphi} + |H(f)|^2 \left(S_{\varphi_f\varphi_f}^1 + \frac{2h\nu}{\eta \bar{P}}\right)}{|1+G(f)H(f)|^2}
\end{equation}
We see that the original phase noise is suppressed, but new noise is added due to detection and fiber noise in setup 1 ($S_{\varphi_f\varphi_f}^1$). This directly allows us to identify the out-of-loop detector (setup 2) noise $S_{\varphi\varphi}^{\mathrm{fb,meas}}$ by substituting $S_{\varphi\varphi}^{\mathrm{fb}}$ into Eq. \ref{eq:phimeas} with uncorrelated fiber and shot noise.

For the in-loop detector (setup 1), however, the same detection noise becomes present in the feedback, which leads to interference. We obtain the following measured noise: 
\begin{multline}
S_{\varphi\varphi}^\mathrm{1,fb,meas} = \frac{S_{\varphi\varphi}}{|1+G(f)H(f)|^2} + \\ + \left[\frac{|H(f)|^2}{|1+G(f)H(f)|^2} + \frac{1}{|G(f)|^2} \left(1 - 2 \mathrm{Re} \frac{G(f)H(f)}{1+G(f)H(f)}\right)\right]\left( S_{\varphi_f\varphi_f}^1 + \frac{2h\nu}{\eta \bar{P}} \right),
\end{multline}
where the first term is the expected reduced laser noise, while the second term represents the detection noise itself as well as its self-interference. In particular, the measured noise in this case may fall well below the original detection noise.
 
In practice, for feedback we employ the I/Q modulation/demodulation technique as described in Sec. \ref{sec:setup}. We can then use the following controller gain model:
\begin{equation}
H(f) = g e^{-2\pi i \tau_\mathrm{D} f + i \vartheta} \frac{\gamma f}{f_\mathrm{c}^2-f^2+i \gamma f},
\end{equation}
where we control the following parameters: gain $g$, phase $\vartheta$, central frequency $f_\mathrm{c}$ and bandwidth $\gamma$.

\section{Experimental results}
\begin{figure}[h!]
	\centering\includegraphics[width=1\linewidth]{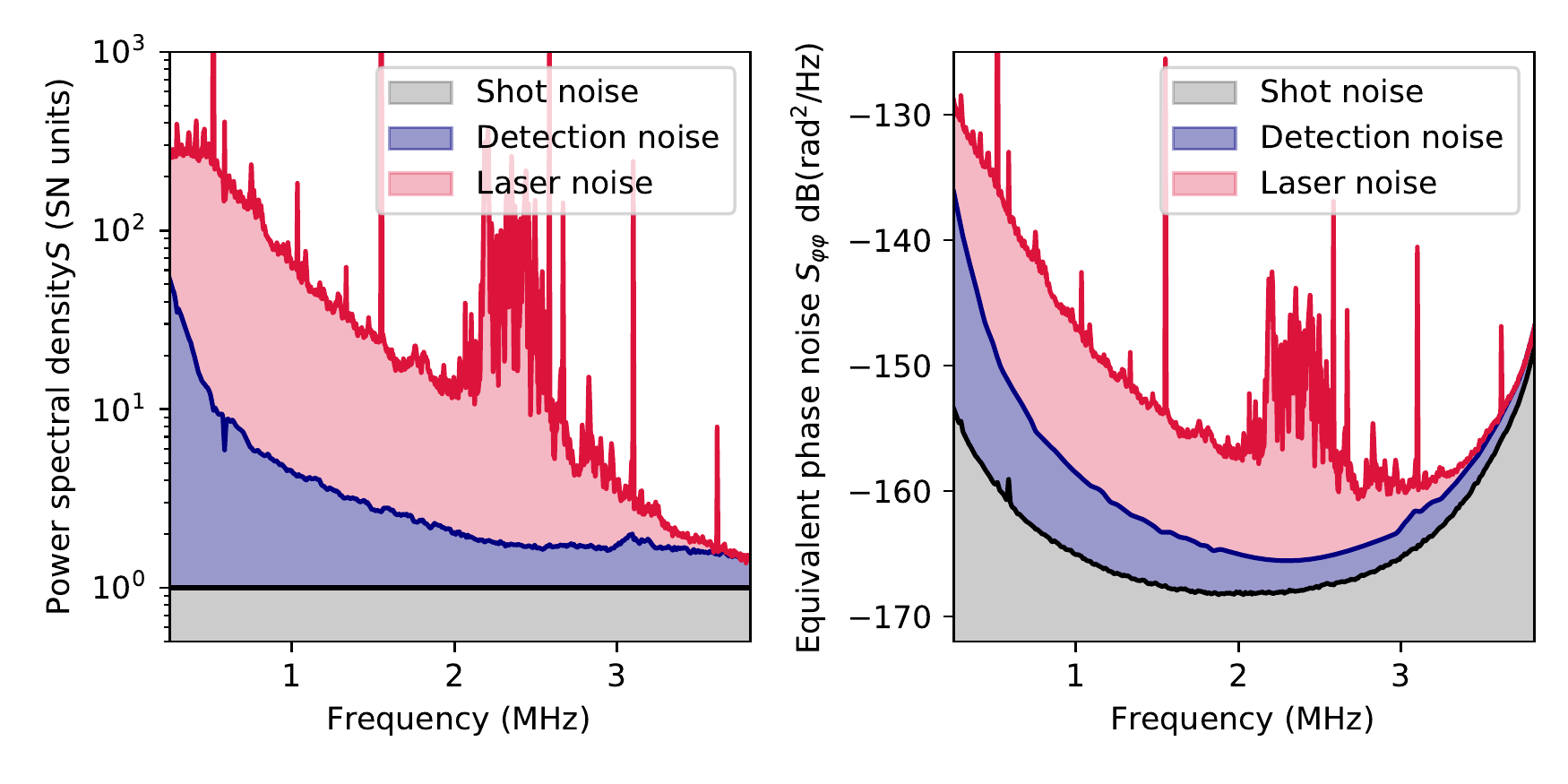}
	\caption{Components of noise registered by the balanced detector in setup 1. (a) Optical power spectral density normalized to photon shot noise and (b) the same signal in terms of equivalent laser phase noise as calculated from transduction of phase noise into detected noise. The three components are photon shot noise, detection noise (primarily composed of fiber noise) and the signal which corresponds to laser phase noise. The peak sensitivity due to delay loop transduction is obtained in the vicinity of \SI{2}{MHz}.}
	\label{fig:master_noise}
	
\end{figure}

\begin{figure}[h!]
	\centering\includegraphics[width=1\linewidth]{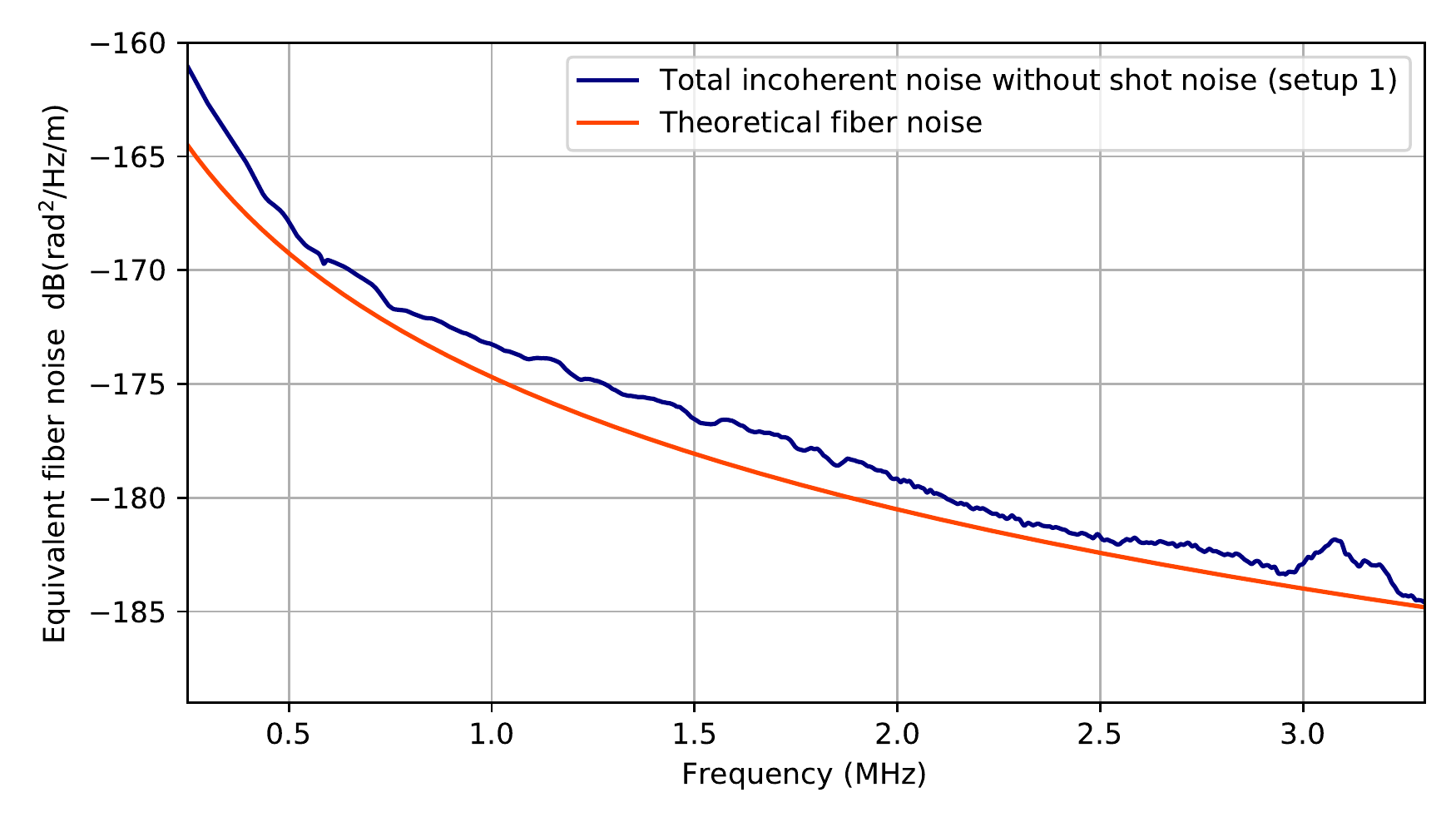}
	\caption{Incoherent part of noise in setup 1 with shot noise subtracted (blue part in \ref{fig:master_noise}) compared with theoretical prediction for noise of the delay fiber. The result is normalized by the equivalent fiber noise per unit fiber length, where we assume that all measured incoherent noise is fiber noise}
\label{fig:fnoise}
\end{figure}

\begin{figure}[h!]
	\centering\includegraphics[width=1\linewidth]{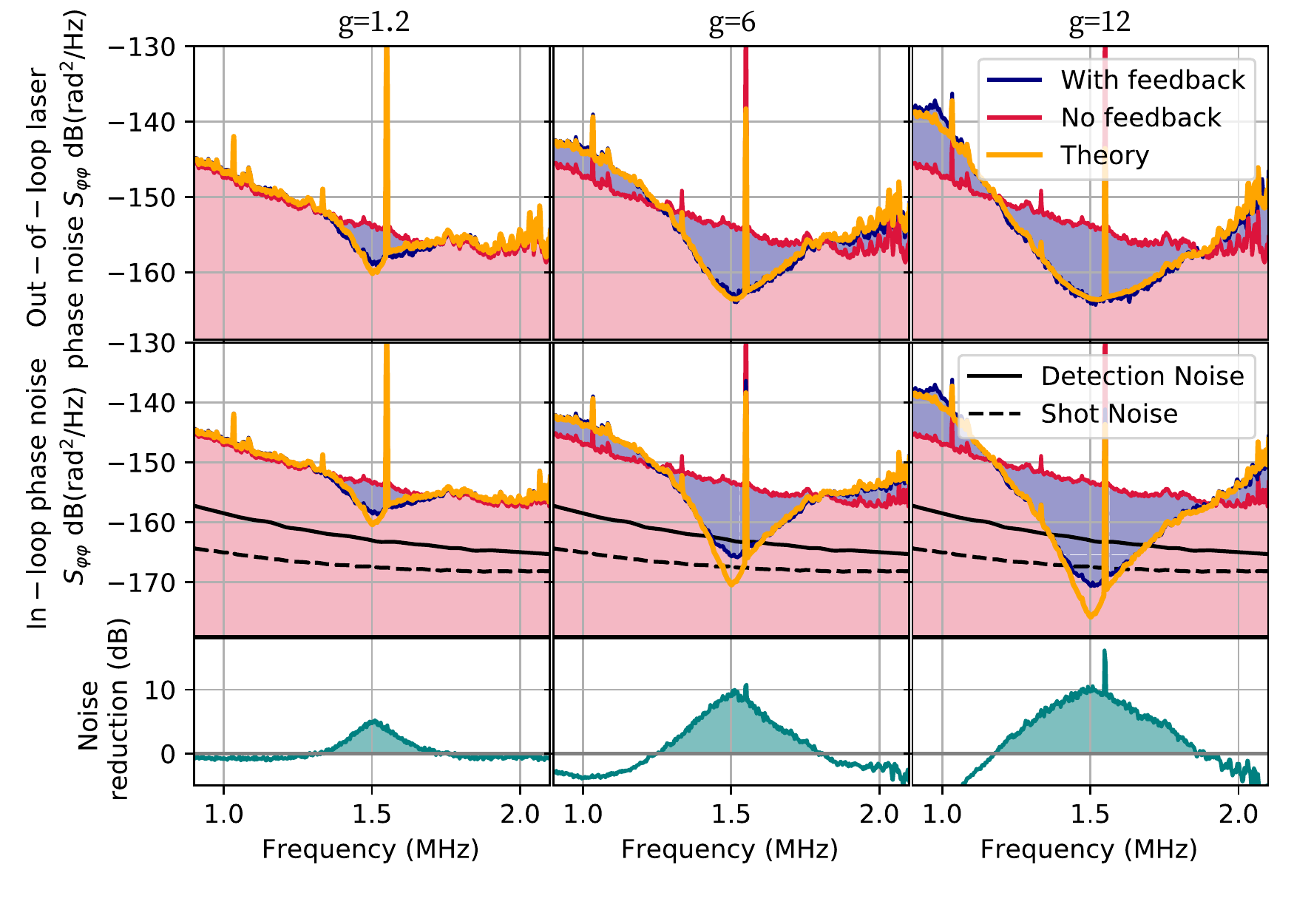}
	\caption{Laser phase noise for loop parameters $f_\mathrm{c}=\SI{1.5}{\mega\hertz}$ and bandwidth $\gamma=\SI{78}{\kilo\hertz}$ and $g=1.2,\ 6$ and $12$ for columns from left to right, respectively. The first row presents the laser phase noise as measured by out-of-loop setup 2, with detection noise subtracted. In the second row we show the in-loop noise of setup 1, converted to equivalent laser phase noise units. Bottom row shows the reduction of laser phase noise, as inferred from the out-of-loop measurement.}
	\label{fig:red15}
	
\end{figure}

\begin{figure}[h!]
	\centering\includegraphics[width=1\linewidth]{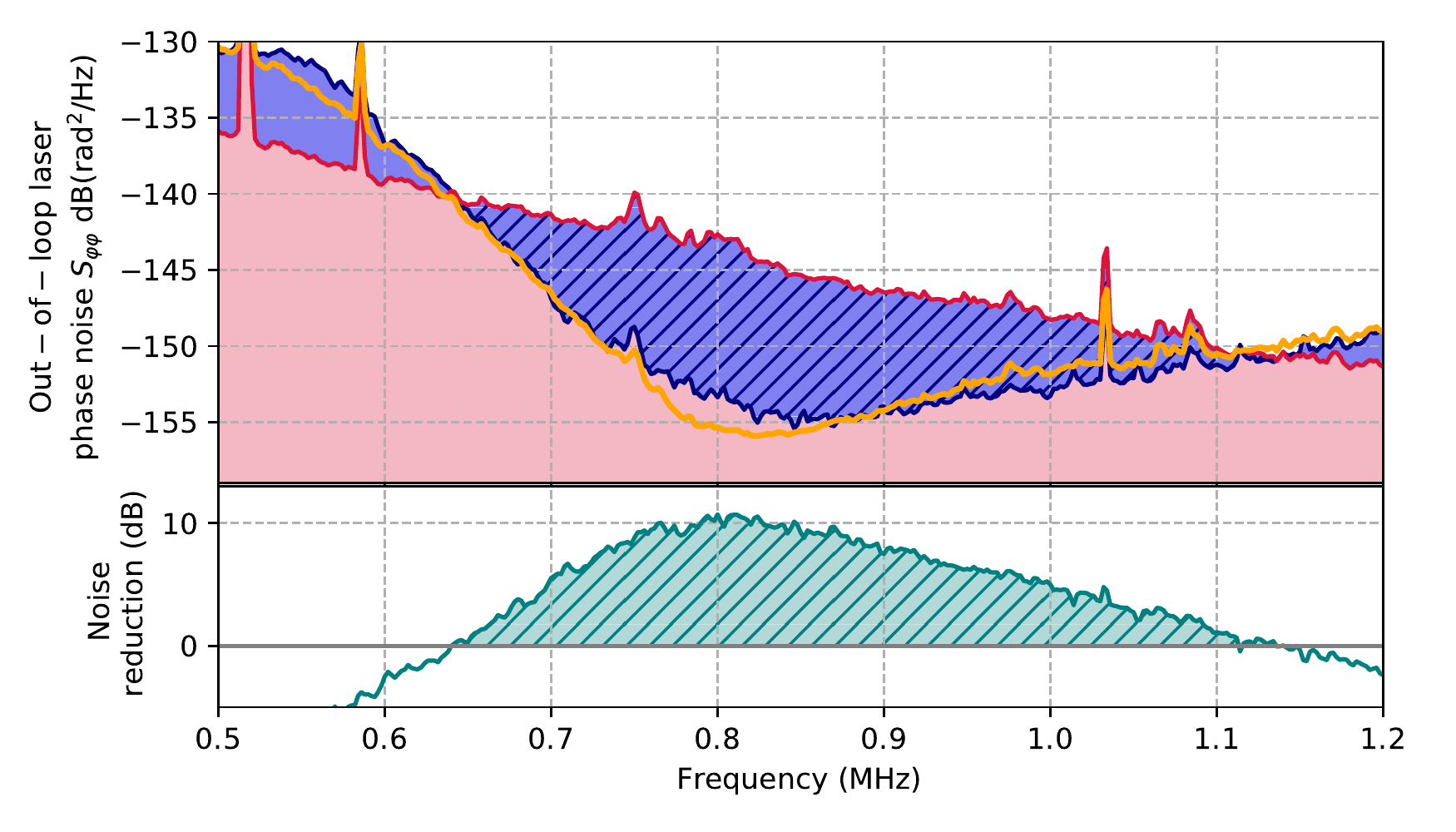}
	\caption{Laser phase noise as measured by out-of-loop setup 2 and respective noise reduction due to feedback with loop parameters: $f_\mathrm{c}=\SI{0.8}{\mega\hertz}$, $\gamma=\SI{78}{\kilo\hertz}$ and $g=10$. Red curve is the original noise, blue represents the noise after feedback, and yellow curve is the theoretical prediction.}
	\label{fig:red08}
	
\end{figure}

\begin{figure}[h!]
	\centering\includegraphics[width=1\linewidth]{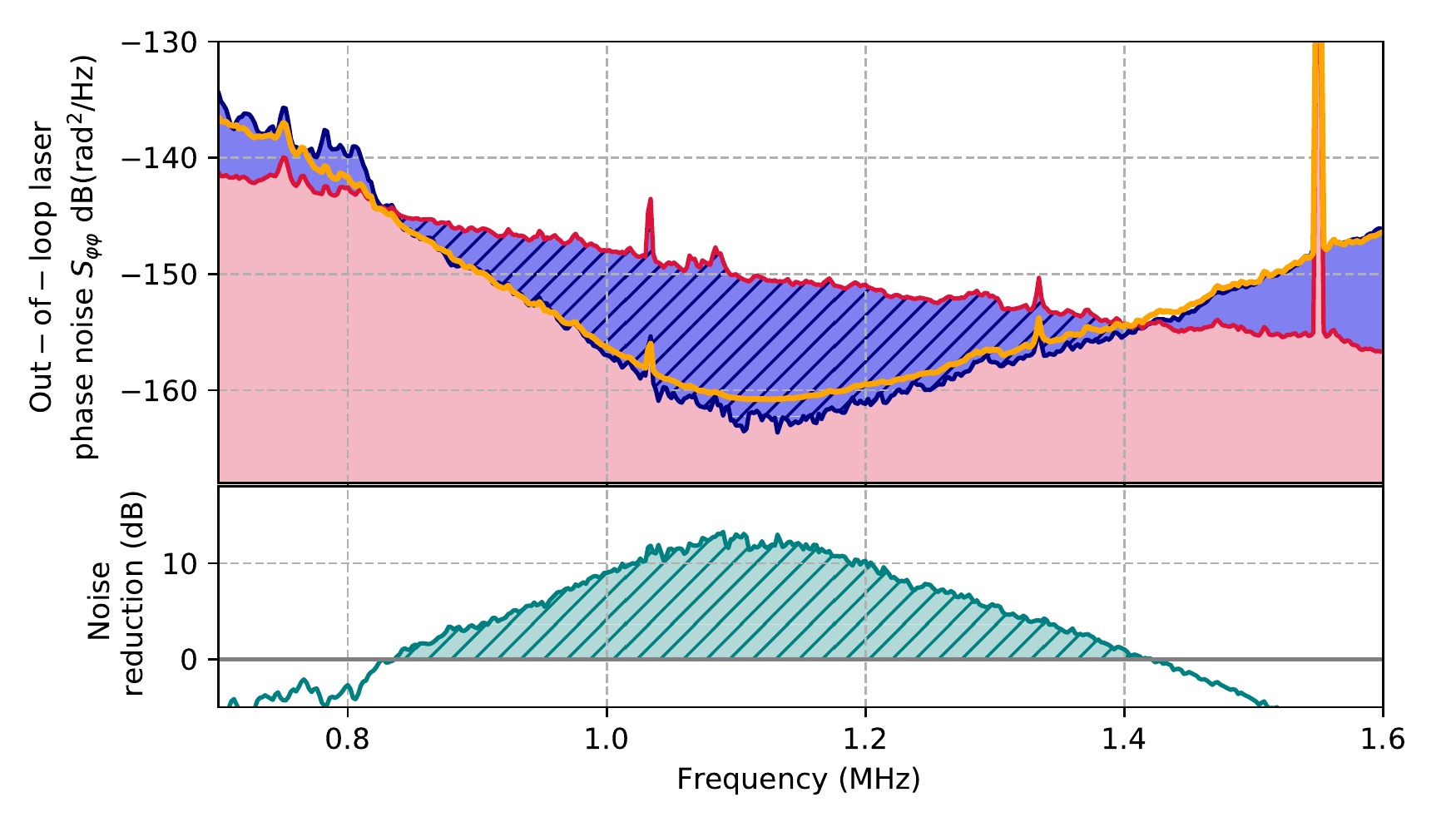}
	\caption{Laser phase noise as measured by out-of-loop setup 2 and respective noise reduction due to feedback with loop parameters: $f_\mathrm{c}=\SI{1.1}{\mega\hertz}$, $\gamma=\SI{78}{\kilo\hertz}$ and $g=6$. Red curve is the original noise, blue represents the noise after feedback, and yellow curve is the theoretical prediction.}
	\label{fig:red11}
	
\end{figure}
In the experiment we use an additional fiber EOM with $V_\pi=\SI{4.65}{\volt}$ to recover an experimental profile of phase sensitivity curve of setup 2. Simultaneously, we may apply a calibration signal to the free-space EOM, which will be detected by both setups, and obtain a relative calibration of setup 1 with respect to calibrated setup 2. For the calibration we apply a low-frequency square-wave pattern and compare amplitudes of its Fourier components as registered by both setups.

Using such calibration, we can determine the cross-spectral density (CSD) which only contains the common-mode laser phase noise (Eq. \ref{eq:csd}). After expressing all PSDs and the CSD in equivalent laser phase noise units (using calibrated phase-modulation responses), we identify the incoherent (uncorrelated) parts of noise in both setups. Finally, we also measure the photon shot noise by subsequently blocking arms of the interferometer and adding resulting registered noise. 

Figure \ref{fig:master_noise} presents the raw measured detector noise (expressed in shot-noise units) and equivalent laser phase noise for setup 1. We decompose the noise into the \emph{coherent} (correlated) laser phase noise and the \emph{incoherent} detection noise part, which is itself composed of shot noise, fiber noise, and other detection noise. While we expect that the incoherent noise is heavily dominated by fiber noise, our analysis does not require this assumption. With $\tau=\SI{250}{\nano\second}$ ($n=1.5$, $L=\SI{50}{\meter}$) we observe maximum sensitivity to phase noise at a \SI{2}{\MHz} offset, and minima of sensitivity at \SI{0}{\MHz} and \SI{4}{MHz}. In terms of equivalent laser phase noise, this means that in the range above \SI{3.5}{\MHz}, where laser phase noise is becoming smaller, we are heavily dominated by detection noise. Below this frequency, however, our detection setup exhibits a signal-to-noise ratio of at least 1. By using an optical power of $\bar{P}=\SI{11}{\milli\watt}$, we make sure that at our main frequency of interest at \SI{1.5}{\MHz}, detection noise is dominated by fiber noise and not shot noise. In particular, fiber noise constitutes around 70\% of detection noise under these conditions at \SI{1.5}{\MHz}. Large laser phase noise peaks around \SI{2.3}{\MHz} make the subtraction procedure sensitive to uncertainties, and thus in order to avoid imprecision we interpolate the detection noise by a second order polynomial in the \SIrange{2}{2.5}{\MHz} region. We have verified that this additional noise is internal to the laser, and most likely originates in the green pump.

With the extracted incoherent noise of setup 1, we may also compare it with a prediction for fiber noise from Eq. \ref{eq:fn}. The following parameters are assumed \cite{PhysRevA.86.023817,6171816}: $T=\SI{298}{\kelvin}$, $\kappa=\SI{1.37}{\watt\per\meter\per\kelvin}$, $\frac{\mathrm{d}n}{\mathrm{d}T}=\SI{1e-5}{\per\kelvin}$, $\alpha=\SI{5e-7}{\per\kelvin}$, $D=\SI{0.82e-6}{\meter^2\per\second}$ and  $r_0=\SI{1.5}{\micro\meter}$. In Fig. \ref{fig:fnoise} observe that while the general behavior is well reproduced, the predicted fiber noise lies slightly below our measured incoherent detection noise (with shot noise subtracted). We attribute this discrepancy either to additional noise, or more likely to inaccuracy in used parameters, which are not directly available for the PM-780HP fiber we use, but are rather extracted from general material properties. In particular, some critical parameters of the model, such as for instance the coefficient of linear expansion vary appreciably across literature references and may also vary depending on the exact fiber material. Furthermore, a strong dependence of the model on fiber core radius as well as temperature call for a more systematic and precise study of fiber noise at MHz frequencies, which to the best of our knowledge has not been performed so far.

Next, we proceed to apply feedback and observe the noise registered in both setups. With other parameters optimized, we use a filter with $f_\mathrm{c}=\SI{1.5}{\mega\hertz}$ and bandwidth $\gamma=\SI{78}{\kilo\hertz}$ while changing the loop gain from $g=1.2$ through $g=6$ to $g=12$. In the top panel of figure \ref{fig:red15} we show the laser phase noise as measured with the out-of-loop setup 2. Here we subtract the uncorrelated incoherent noise. We observe that with our feedback we register significantly less noise with up to \SI{10}{\dB} of reduction for the highest gain (see the bottom row). A phase noise peak at \SI{1.54}{\MHz} is suppressed to an even greater degree. Simultaneously, we observe increased noise away from the central frequency, which becomes more pronounced at higher gains. This is due to sub-optimal phase of the feedback controller at those frequencies. For the purpose of noise reduction at the specific frequencies of our optomechanical experiment, this does not pose a problem, as long as no oscillation or feedback instability is observed.

In the middle row we show the simultaneously registered noise of the in-loop setup 1. This noise is the raw noise of the detector, converted into equivalent laser phase noise units. Here we observe that the noise, particularly for the highest gain, reaches well below the detection noise. This noise squashing behavior is expected as the result of noise self-interference. Our experimental results are accompanied by theory curves that use the model from Sec. \ref{sec:model} along with measured responses and detection noise. We observe particularly good agreement for the most important phase noise measured by the out-of-loop detector. A slightly higher noise than expected is registered in the in-loop detector, which may be a result of additional electronic detection noise.

Finally, we apply the feedback at different central frequencies, as shown in Figs. \ref{fig:red08} and \ref{fig:red11}. We observe up to \SI{12}{\dB} of noise reduction and bandwidths of at least \SI{300}{\kHz} in all cases.

\section{Conclusions and prospects}
We have demonstrated a setup that combines a measurement of laser phase noise via a delay line with an electro-optic feedback to actively reduce laser phase noise in the \si{\MHz} frequency range, which lies outside of previously explored frequency domains \cite{Li:17,Chen:89,Cranch:02,Sheard:06,Kefelian:09,Dong:15,  _m_d_2015,Shehzad:19,doi:10.1063/1.3606439,Bandutunga:20}. We have demonstrated phase noise reduction of at least \SI{10}{\dB} and broadband operation, achieving relatively low phase noise densities at the level of approximately \SI{-160}{\deci\bel(\radian^2/\hertz)}, or equivalently \SI{-20}{\deci\bel((\radian\per\second)^2/\hertz)} in terms of frequency noise. We have also demonstrated an effective model for predicting the performance of our method, as well as a convenient way to identify noise components in the system based on cross-spectral density evaluation. Our results also include measurements of the fiber noise at high frequencies, which show reasonable agreement with predicted thermoconductive noise.

 Our results can find particular applications in optomechanics, where reduction of noise around a specific offset frequency is desired. This applies to both sideband and feedback (cold-damping) cooling of trapped particle oscillators \cite{Delic892,PhysRevLett.122.123602,PhysRevLett.123.153601}, room-temperature integrated resonators \cite{PhysRevLett.123.223602}, membrane-in-the-middle systems \cite{ourOptica,Rossi2018} and others \cite{Kippenberg_2013,Safavi_Naeini_2013}.
Other applications sensitive to noise at a particular offset frequency include the driving of Raman transitions \cite{Arias:17,LeGouet2008} or hybrid electro-opto-mechanical converters \cite{Bagci2014}.

Our main prospective application is generation of low-phase-noise light for optomechanical sideband cooling. As estimated by Kippenberg \emph{et al.} \cite{Kippenberg_2013}, the limit for the expected final phonon occupation (i.e. with intracavity power and detuning optimized) for an optomechanical resonator in the sideband-resolved regime due to phase noise is given approximately by:
\begin{equation}
\bar{n}\approx\sqrt{\frac{\bar{n}_\mathrm{th} \Gamma_\mathrm{m}}{g_0^2} \Omega_\mathrm{m}^2 \bar{S}_{\varphi\varphi}(\Omega_\mathrm{m})}
\end{equation}
Assuming a set of parameters consistent with recent experiments \cite{ourOptica} ($g_0/2\pi=\SI{8}{\hertz}$, $\bar{n}_\mathrm{th}/2\pi=1.5\times10^5$, $\Gamma_\mathrm{m}/2\pi=\SI{2}{\milli\hertz}$ and $\Omega_\mathrm{m}/2\pi=\SI{1.5}{\mega\hertz}$), we obtain $\bar{n}=0.16$ without our noise cancellation device ($\bar{S}_{\varphi\varphi}=\SI{-154}{\deci\bel(\radian^2/\hertz)}$) which is improved to $\bar{n}=0.05$ with reduced laser phase noise ($\bar{S}_{\varphi\varphi}=\SI{-164}{\deci\bel(\radian^2/\hertz)}$). Even more remarkably, at around $\Omega_\mathrm{m}/2\pi=\SI{1}{\mega\hertz}$ we may decrease residual occupation from $\bar{n}=0.24$ to $\bar{n}=0.05$ as well.

It is also worth mentioning an alternative approach to reduce noise at \si{MHz} frequencies. This approach makes use of a filtering cavity, which has been also demonstrated as a method to measure laser phase noise \cite{Kippenberg_2013}. In such setup, one uses light directly transmitted through a narrowband cavity. However, special treatment would be required to make sure that mirror thermal noise, which is widely considered an important limitation of optomechanical setups \cite{Zhao:12}, remains well below the input laser phase noise. We estimate that such cavity would either have to be sufficiently long (we estimate at least 30 cm) to reduce mirror noise transduction, or cryogenically cooled to suppress thermal motion. Therefore, a fiber delay line seems to be a more practical solution in several matters. Another advantage over a filtering cavity is that it requires only a constant power, and does not produce a loss for main experimental light.

Finally, we envisage that the delay line approach can be further improved by reducing fiber thermal noise. One approach would be to embed the delay line in a cryogenic environment. We estimate that due to simultaneous reduction of thermal noise and the thermorefractive coefficient, the fiber noise could be reduced by more than \SI{20}{\dB} at moderate liquid-nitrogen temperatures. Furthermore, the noise can be greatly reduced with an increased mode field diameter. For a given wavelength, this could be possibly achieved in large-mode-area fibers.

\section*{Funding}
Villum Fonden (QMACVillum Investigator Grant 25880); European Research Council (Advanced Grant QUANTUM-N); John Templeton Foundation;

\section*{Acknowledgments}
We thank J. H. Müller, Y. Tsaturyan and G. Enzian for discussions and input.
M.P. is also supported by the MAB/2018/4 project "Quantum Optical Technologies", carried out within the International Research Agendas program of the Foundation for Polish Science co-financed by the European Union under the European Regional Development Fund. M.P. and I.G. contributed equally to this work. E.S.P. is a Villum Investigator.

\section*{Disclosures}
The authors declare no conflicts of interest.

%%%%%%%%%%%%%%%%%%%%%%% References %%%%%%%%%%%%%%%%%%%%%%%%%

\bibliography{biblio}

\begin{thebibliography}{10}
\newcommand{\enquote}[1]{``#1''}

\bibitem{PhysRev.112.1940}
A.~L. Schawlow and C.~H. Townes, \enquote{Infrared and optical masers,}
  {\protect\JournalTitle{Phys. Rev.}} \textbf{112}, 1940--1949 (1958).

\bibitem{NUMATA2012798}
K.~Numata and J.~Camp, \enquote{Estimation of frequency noise in semiconductor
  lasers due to mechanical thermal noise,} {\protect\JournalTitle{Physics
  Letters A}} \textbf{376}, 798 -- 802 (2012).

\bibitem{CONFORTI2020126286}
E.~Conforti, M.~Rodigheri, T.~Sutili, and F.~J. Galdieri, \enquote{Acoustical
  and $1/f$ noises in narrow linewidth lasers,} {\protect\JournalTitle{Optics
  Communications}} \textbf{476}, 126286 (2020).

\bibitem{Kessler2012}
T.~Kessler, C.~Hagemann, C.~Grebing, T.~Legero, U.~Sterr, F.~Riehle, M.~J.
  Martin, L.~Chen, and J.~Ye, \enquote{A sub-40-mhz-linewidth laser based on a
  silicon single-crystal optical cavity,} {\protect\JournalTitle{Nature
  Photonics}} \textbf{6}, 687--692 (2012).

\bibitem{PhysRevA.77.053809}
J.~Alnis, A.~Matveev, N.~Kolachevsky, T.~Udem, and T.~W. H\"ansch,
  \enquote{Subhertz linewidth diode lasers by stabilization to vibrationally
  and thermally compensated ultralow-expansion glass fabry-p\'erot cavities,}
  {\protect\JournalTitle{Phys. Rev. A}} \textbf{77}, 053809 (2008).

\bibitem{8123620}
M.~{Tawfieq}, A.~K. {Hansen}, O.~B. {Jensen}, D.~{Marti}, B.~{Sumpf}, and P.~E.
  {Andersen}, \enquote{Intensity noise transfer through a diode-pumped titanium
  sapphire laser system,} {\protect\JournalTitle{IEEE Journal of Quantum
  Electronics}} \textbf{54}, 1700209 (2018).

\bibitem{LUDVIGSEN1998180}
H.~Ludvigsen, M.~Tossavainen, and M.~Kaivola, \enquote{Laser linewidth
  measurements using self-homodyne detection with short delay,}
  {\protect\JournalTitle{Optics Communications}} \textbf{155}, 180 -- 186
  (1998).

\bibitem{Li:17}
D.~Li, C.~Qian, Y.~Li, and J.~Zhao, \enquote{Efficient laser noise reduction
  method via actively stabilized optical delay line,}
  {\protect\JournalTitle{Opt. Express}} \textbf{25}, 9071--9077 (2017).

\bibitem{Chen:89}
Y.~T. Chen, \enquote{Use of single-mode optical fiber in the stabilization of
  laser frequency,} {\protect\JournalTitle{Appl. Opt.}} \textbf{28}, 2017--2021
  (1989).

\bibitem{Cranch:02}
G.~A. Cranch, \enquote{Frequency noise reduction in erbium-doped fiber
  distributed-feedback lasers by electronic feedback,}
  {\protect\JournalTitle{Opt. Lett.}} \textbf{27}, 1114--1116 (2002).

\bibitem{Sheard:06}
B.~S. Sheard, M.~B. Gray, and D.~E. McClelland, \enquote{High-bandwidth laser
  frequency stabilization to a fiber-optic delay line,}
  {\protect\JournalTitle{Appl. Opt.}} \textbf{45}, 8491--8499 (2006).

\bibitem{Kefelian:09}
F.~K\'{e}f\'{e}lian, H.~Jiang, P.~Lemonde, and G.~Santarelli,
  \enquote{Ultralow-frequency-noise stabilization of a laser by locking to an
  optical fiber-delay line,} {\protect\JournalTitle{Opt. Lett.}} \textbf{34},
  914--916 (2009).

\bibitem{Dong:15}
J.~Dong, Y.~Hu, J.~Huang, M.~Ye, Q.~Qu, T.~Li, and L.~Liu, \enquote{Subhertz
  linewidth laser by locking to a fiber delay line,}
  {\protect\JournalTitle{Appl. Opt.}} \textbf{54}, 1152--1156 (2015).

\bibitem{_m_d_2015}
R.~Šmíd, M.~Čížek, B.~Mikel, and O.~Číp, \enquote{Frequency noise
  suppression of a single mode laser with an unbalanced fiber interferometer
  for subnanometer interferometry,} {\protect\JournalTitle{Sensors}}
  \textbf{15}, 1342–1353 (2015).

\bibitem{Shehzad:19}
A.~Shehzad, P.~Brochard, R.~Matthey, T.~S\"{u}dmeyer, and S.~Schilt,
  \enquote{10 khz linewidth mid-infrared quantum cascade laser by stabilization
  to an optical delay line,} {\protect\JournalTitle{Opt. Lett.}} \textbf{44},
  3470--3473 (2019).

\bibitem{doi:10.1063/1.3606439}
W.-K. Lee, C.~Y. Park, J.~Mun, and D.-H. Yu, \enquote{Linewidth reduction of a
  distributed-feedback diode laser using an all-fiber interferometer with short
  path imbalance,} {\protect\JournalTitle{Review of Scientific Instruments}}
  \textbf{82}, 073105 (2011).

\bibitem{Bandutunga:20}
C.~P. Bandutunga, T.~G. McRae, Y.~Zhang, M.~B. Gray, and J.~H. Chow,
  \enquote{Infrasonic performance of a passively stabilized, all-fiber, optical
  frequency reference,} {\protect\JournalTitle{Opt. Express}} \textbf{28},
  9280--9287 (2020).

\bibitem{PhysRevA.86.023817}
L.~Duan, \enquote{General treatment of the thermal noises in optical fibers,}
  {\protect\JournalTitle{Phys. Rev. A}} \textbf{86}, 023817 (2012).

\bibitem{Ma:94}
L.-S. Ma, P.~Jungner, J.~Ye, and J.~L. Hall, \enquote{Delivering the same
  optical frequency at two places: accurate cancellation of phase noise
  introduced by an optical fiber or other time-varying path,}
  {\protect\JournalTitle{Opt. Lett.}} \textbf{19}, 1777--1779 (1994).

\bibitem{PhysRevLett.111.110801}
S.~Droste, F.~Ozimek, T.~Udem, K.~Predehl, T.~W. H\"ansch, H.~Schnatz,
  G.~Grosche, and R.~Holzwarth, \enquote{Optical-frequency transfer over a
  single-span 1840 km fiber link,} {\protect\JournalTitle{Phys. Rev. Lett.}}
  \textbf{111}, 110801 (2013).

\bibitem{doi:10.1063/1.4939918}
J.~Dong, J.~Huang, T.~Li, and L.~Liu, \enquote{Observation of fundamental
  thermal noise in optical fibers down to infrasonic frequencies,}
  {\protect\JournalTitle{Applied Physics Letters}} \textbf{108}, 021108 (2016).

\bibitem{Gagliardi1081}
G.~Gagliardi, M.~Salza, S.~Avino, P.~Ferraro, and P.~De~Natale,
  \enquote{Probing the ultimate limit of fiber-optic strain sensing,}
  {\protect\JournalTitle{Science}} \textbf{330}, 1081--1084 (2010).

\bibitem{Cranch286}
G.~A. Cranch and S.~Foster, \enquote{Comment on {\textquotedblleft}probing the
  ultimate limit of fiber-optic strain sensing{\textquotedblright},}
  {\protect\JournalTitle{Science}} \textbf{335}, 286--286 (2012).

\bibitem{Tsaturyan2017}
Y.~Tsaturyan, A.~Barg, E.~S. Polzik, and A.~Schliesser, \enquote{Ultracoherent
  nanomechanical resonators via soft clamping and dissipation dilution,}
  {\protect\JournalTitle{Nature Nanotechnology}} \textbf{12}, 776--783 (2017).

\bibitem{ourOptica}
I.~Galinskiy, Y.~Tsaturyan, M.~Parniak, and E.~S. Polzik, \enquote{Phonon
  counting thermometry of an ultracoherent membrane resonator near its motional
  ground state,} {\protect\JournalTitle{Optica}} \textbf{7}, 718--725 (2020).

\bibitem{PhysRevLett.123.153601}
N.~Meyer, A.~d. l.~R. Sommer, P.~Mestres, J.~Gieseler, V.~Jain, L.~Novotny, and
  R.~Quidant, \enquote{Resolved-sideband cooling of a levitated nanoparticle in
  the presence of laser phase noise,} {\protect\JournalTitle{Phys. Rev. Lett.}}
  \textbf{123}, 153601 (2019).

\bibitem{Kippenberg_2013}
T.~J. Kippenberg, A.~Schliesser, and M.~L. Gorodetsky, \enquote{Phase noise
  measurement of external cavity diode lasers and implications for
  optomechanical sideband cooling of {GHz} mechanical modes,}
  {\protect\JournalTitle{New Journal of Physics}} \textbf{15}, 015019 (2013).

\bibitem{Safavi_Naeini_2013}
A.~H. Safavi-Naeini, J.~Chan, J.~T. Hill, S.~Gröblacher, H.~Miao, Y.~Chen,
  M.~Aspelmeyer, and O.~Painter, \enquote{Laser noise in cavity-optomechanical
  cooling and thermometry,} {\protect\JournalTitle{New Journal of Physics}}
  \textbf{15}, 035007 (2013).

\bibitem{Rossi2018}
M.~Rossi, D.~Mason, J.~Chen, Y.~Tsaturyan, and A.~Schliesser,
  \enquote{Measurement-based quantum control of mechanical motion,}
  {\protect\JournalTitle{Nature}} \textbf{563}, 53--58 (2018).

\bibitem{TheJurgen}
J.~Appel, D.~Hoffman, E.~Figueroa, and A.~I. Lvovsky, \enquote{Electronic noise
  in optical homodyne tomography,} {\protect\JournalTitle{Phys. Rev. A}}
  \textbf{75}, 035802 (2007).

\bibitem{8087380}
L.~{Neuhaus}, R.~{Metzdorff}, S.~{Chua}, T.~{Jacqmin}, T.~{Briant},
  A.~{Heidmann}, P.-F. {Cohadon}, and S.~{Deléglise}, \enquote{Pyrpl (python
  red pitaya lockbox) — an open-source software package for fpga-controlled
  quantum optics experiments,} in \emph{2017 Conference on Lasers and
  Electro-Optics Europe European Quantum Electronics Conference
  (CLEO/Europe-EQEC),}  (2017).

\bibitem{GorodetksyDOVOCRUFNC}
M.~L. Gorodetsky, A.~Schliesser, G.~Anetsberger, S.~Deleglise, and T.~J.
  Kippenberg, \enquote{Determination of the vacuum optomechanical coupling rate
  using frequency noise calibration,} {\protect\JournalTitle{Opt. Express}}
  \textbf{18}, 23236--23246 (2010).

\bibitem{6171816}
R.~E. {Bartolo}, A.~B. {Tveten}, and A.~{Dandridge}, \enquote{Thermal phase
  noise measurements in optical fiber interferometers,}
  {\protect\JournalTitle{IEEE Journal of Quantum Electronics}} \textbf{48},
  720--727 (2012).

\bibitem{Delic892}
U.~Deli{\'c}, M.~Reisenbauer, K.~Dare, D.~Grass, V.~Vuleti{\'c}, N.~Kiesel, and
  M.~Aspelmeyer, \enquote{Cooling of a levitated nanoparticle to the motional
  quantum ground state,} {\protect\JournalTitle{Science}} \textbf{367},
  892--895 (2020).

\bibitem{PhysRevLett.122.123602}
U.~Deli{\'c}, M.~Reisenbauer, D.~Grass, N.~Kiesel,
  V.~Vuleti\ifmmode~\acute{c}\else \'{c}\fi{}, and M.~Aspelmeyer,
  \enquote{Cavity cooling of a levitated nanosphere by coherent scattering,}
  {\protect\JournalTitle{Phys. Rev. Lett.}} \textbf{122}, 123602 (2019).

\bibitem{PhysRevLett.123.223602}
J.~Guo, R.~Norte, and S.~Gr\"oblacher, \enquote{Feedback cooling of a room
  temperature mechanical oscillator close to its motional ground state,}
  {\protect\JournalTitle{Phys. Rev. Lett.}} \textbf{123}, 223602 (2019).

\bibitem{Arias:17}
N.~Arias, V.~Abediyeh, S.~Hamzeloui, and E.~Gomez, \enquote{Low phase noise
  beams for raman transitions with a phase modulator and a highly birefringent
  crystal,} {\protect\JournalTitle{Opt. Express}} \textbf{25}, 5290--5301
  (2017).

\bibitem{LeGouet2008}
J.~Le~Gou{\"e}t, T.~E. Mehlst{\"a}ubler, J.~Kim, S.~Merlet, A.~Clairon,
  A.~Landragin, and F.~Pereira Dos~Santos, \enquote{Limits to the sensitivity
  of a low noise compact atomic gravimeter,} {\protect\JournalTitle{Applied
  Physics B}} \textbf{92}, 133--144 (2008).

\bibitem{Bagci2014}
T.~Bagci, A.~Simonsen, S.~Schmid, L.~G. Villanueva, E.~Zeuthen, J.~Appel, J.~M.
  Taylor, A.~S{\o}rensen, K.~Usami, A.~Schliesser, and E.~S. Polzik,
  \enquote{Optical detection of radio waves through a nanomechanical
  transducer,} {\protect\JournalTitle{Nature}} \textbf{507}, 81--85 (2014).

\bibitem{Zhao:12}
Y.~Zhao, D.~J. Wilson, K.-K. Ni, and H.~J. Kimble, \enquote{Suppression of
  extraneous thermal noise in cavity optomechanics,}
  {\protect\JournalTitle{Opt. Express}} \textbf{20}, 3586--3612 (2012).

\end{thebibliography}

\end{document}